\documentclass[11pt,twoside]{article}

\usepackage{macroaaa-eng}
\usepackage[english]{babel}
\usepackage{psfig}
\usepackage{graphicx}
\usepackage{array}
\usepackage[T1]{fontenc} 
\usepackage{amsmath}
\usepackage{amssymb}    
\usepackage{latexsym}
\usepackage{verbatim}
\usepackage{natbib}
\usepackage{endnotes}
\long\def\symbolfootnote[#1]#2{\begingroup%
\def\thefootnote{\fnsymbol{footnote}}\footnote[#1]{#2}\endgroup}
\begin{document}

\vskip 1.0cm
\markboth{A.~Zajczyk}{Ambiguity of converting phase-averaged flux into luminosity
for MSPs}
\pagestyle{myheadings}

\vspace*{0.5cm}
\title{Ambiguity of converting phase-averaged flux into luminosity
for millisecond pulsars in gamma rays}

\author{Anna~Zajczyk}
\affil{Nicolaus Copernicus Astronomical Center, ul. Rabia\'{n}ska 8, 87-100 Toru\'{n}, Poland\\
       \vspace{2.2mm}zajczyk@ncac.torun.pl}

\begin{abstract}We study a magnitude of possible over/underestimation of the actual $\gamma$-ray luminosity $L_{\mathrm{actual}}$ of a millisecond pulsar when using so-called  pseudo luminosity $L_{\mathrm{pseudo}}$ which is inferred from a phase-averaged flux. Both, $L_{\mathrm{actual}}$ and $L_{\mathrm{pseudo}}$ are calculated with the numerical code that models the millisecond pulsar magnetospheric activity in the space charge limited flow approximation with unscreened accelerating electric field. The behaviour of $L_{\mathrm{pseudo}}$/$L_{\mathrm{actual}}$ depending on viewing angle is analysed in the energy bands corresponding to the operational energy bands of Fermi GST and H.E.S.S. II.
\end{abstract}

\section{Introduction}

Magnetospheric activity of the millisecond pulsars is numerically modelled in order to test the correctness of pulsar theory and also in order to develop the tool that would predict the detectability of these pulsars with high energy observatories.

Our main reason for carrying out the numerical modelling of radiative processes taking place in millisecond pulsar magnetosphere is to check the accuracy of the so-called pseudo luminosity $L_{\mathrm{pseudo}}$ which is inferred from the observed phase-averaged flux as an estimation of the pulsar luminosity in $\gamma$-rays. In the calculations the unscreened electric field that includes general relativity effects is taken into account. The calculations are made for \textit{a typical} millisecond pulsar with the spin period $P = 2.3$ ms and the magnetic field strength at the pole $B_{\mathrm{pc}} = 10^{-3}$ TG.
We restrict our analysis to the energy range of $\gamma$-rays where the emission is due to curvature process. Analogous treatment of X-ray characteristics of millisecond pulsars would require taking into account synchrotron radiation due to secondary pairs
as well as thermal component and this is beyond the scope of this work.

The main features of the numerical model are presented in Section 2; Sect. 3 shows the preliminary results of the modelling. The concluding remarks are gathered in Sect. 4.

\section{Numerical Model}\label{sec2}

The radiative processes taking place in pulsar magnetosphere are modelled in 3D. It is assumed that charged particles move along open magnetic field lines where their flow is treated within the framework of the space charge limited flow \citep{fas77}. The particles accelerate in the unscreened electric field (Muslimov \& Harding, 1997; Harding \& Muslimov, 1998) that includes the general relativistic effect of dragging of inertial frames of reference \citep{mt92}. It is predicted that for the majority of millisecond pulsars the acceleration voltage drop that develops above polar cap is not limited by the development of a pair formation front \citep{hmz02}. Thus, particles are accelerated even at high altitudes and it is justified to describe the accelerating electric field as being unscreened. 

In the magnetospheres of millisecond pulsars a variety of radiative processes may take place. 
In this work, however, we restrict the considered processes to curvature radiation and the magnetic absorption of curvature photons without tracing the created secondary particles because it is the curvature emission which (according to polar cap
models) dominates completely the $\gamma$-ray energy range in millisecond pulsars \citep{brd00}.
 The escape rate of primary electrons from the neutron star surface is assumed to be the Goldreich-Julian rate in the general relativistic regime \citep{hm98}.

 \begin{figure} 
  \begin{center}
\hspace{0.2cm}
    \psfig{figure=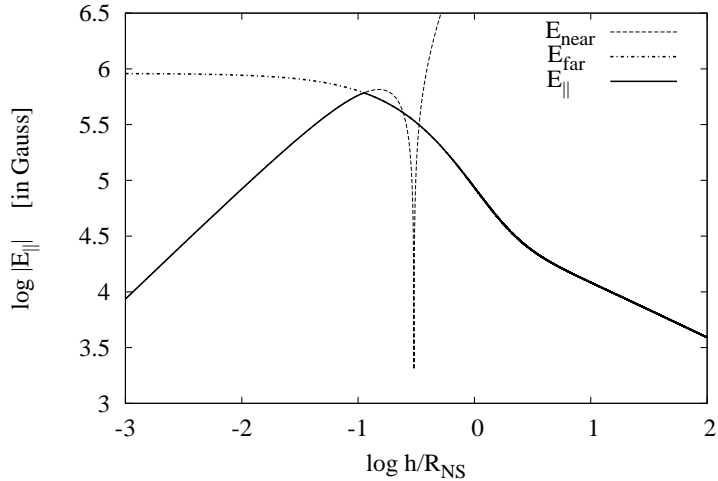,width=10cm}
    \caption{\small The accelerating electric field for a millisecond pulsar characterised with $P=2.3$~ms, $B_{\mathrm{pc}}=10^{-3}$ TG, $\alpha=20^{\circ}$ and $\xi=0.7$. The total electric field $E_{||}$(solid line) is a combination of a formula describing the electric field near the neutron star $E_{\mathrm{near}}$ (dashed line) up to a distance $h \sim 0.7$ of the polar cap radius where a formula depicted with $E_{\mathrm{far}}$ (dot-dashed line) becomes applicable.}\label{acc_tot}
  \end{center}
\end{figure}
 
Figure~\ref{acc_tot} demonstrates the behaviour of the electric field $E_{||}$ - the component parallel to the local magnetic field - above the polar cap (solid line). The accelerating field is a combination of the formula describing the electric field in the vicinity of a pulsar (dashed line; Dyks \& Rudak, 2000) and the formula applicable to distances beyond $\sim 0.7$ of the polar cap radius (dot-dashed line; Muslimov \& Harding, 1997). In Fig.~\ref{acc_tot} the case of pulsar with $P=2.3$~ms, $B_{\mathrm{pc}}=10^{-3}$ TG, $\alpha = 20^{\circ}$ and $\xi = 0.7$ is shown.  Here $\alpha$ is an inclination angle of the magnetic axis with respect to the spin axis and $\xi \equiv \frac{\theta}{\theta(\eta)}$, where $\theta$ is a magnetic colatitude and $\theta(\eta)$ is a half-opening angle of the polar magnetic flux tube; $\eta \equiv 1 + \frac{h}{R_{\mathrm{NS}}}$, where $h$ is an altitude above the neutron star surface.

\section{Results}\label{sec3}

The parameters of the modelled \textit{typical} millisecond pulsar that were used in calculations are: radius $R_{\mathrm{NS}} = 10^{6}$ cm, mass $M_{\mathrm{NS}} = 1.4$ $\mathrm{M}_{\odot}$ , moment of inertia $I_{\mathrm{NS}} = 10^{45}$ g cm$^{2}$, spin period $P = 2.3$ ms, magnetic field strength at the pole $B_{\mathrm{pc}} = 10^{-3}$ TG, inclination angle $\alpha = 20^{\circ}$ and spin down luminosity $L_{\mathrm{sd}} \simeq 4\pi^{4}c^{-3} B_{\mathrm{pc}}^{2}\; R_{\mathrm{NS}}^{6}\; P^{-4} \simeq 5.2 \cdot 10^{35}$ erg s$^{-1}$.

\begin{figure} 
\hspace{0.2cm} \vbox{
    \includegraphics[scale=0.7]{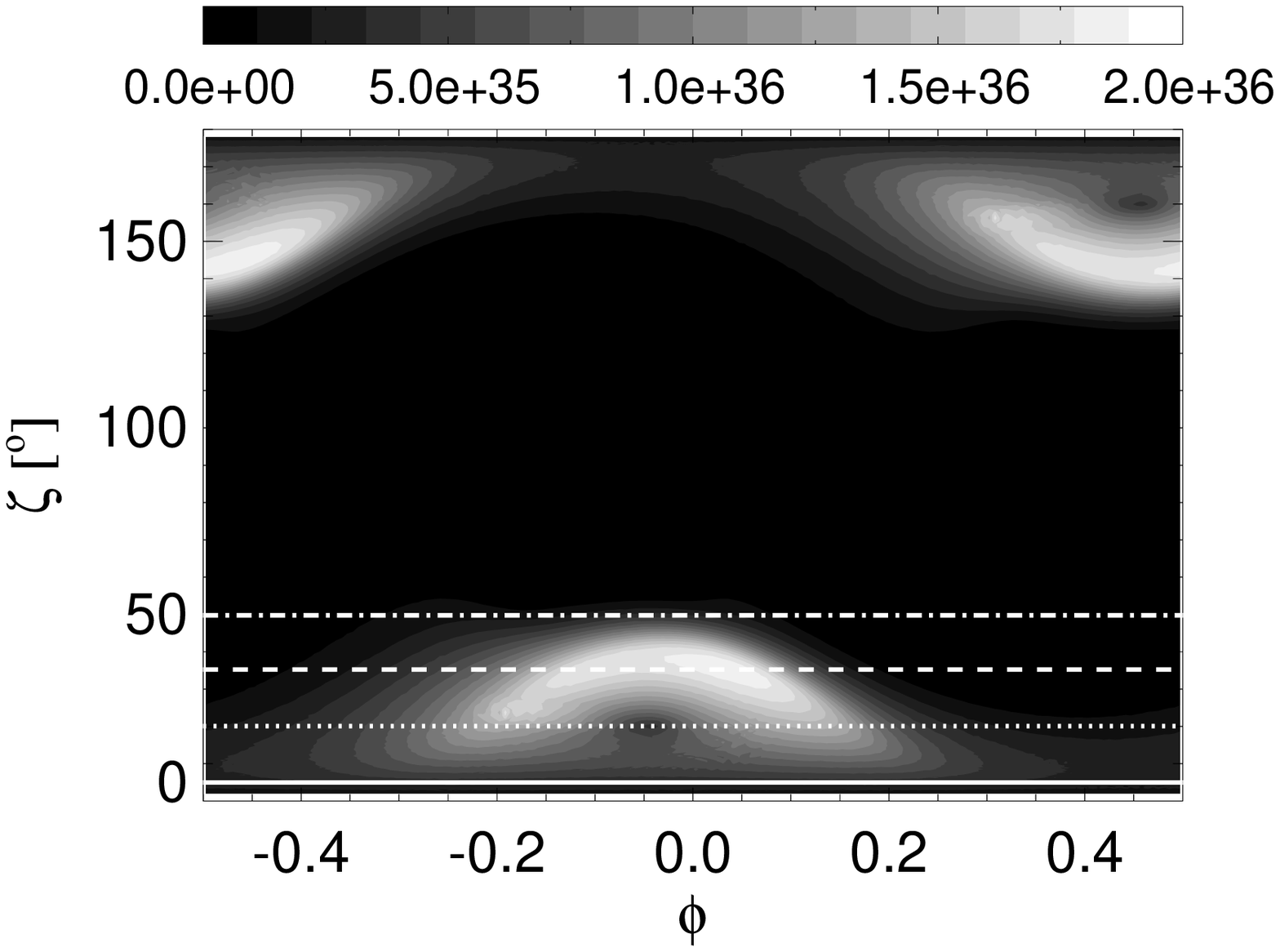} 

    \vspace{1.0cm} \includegraphics[scale=0.7]{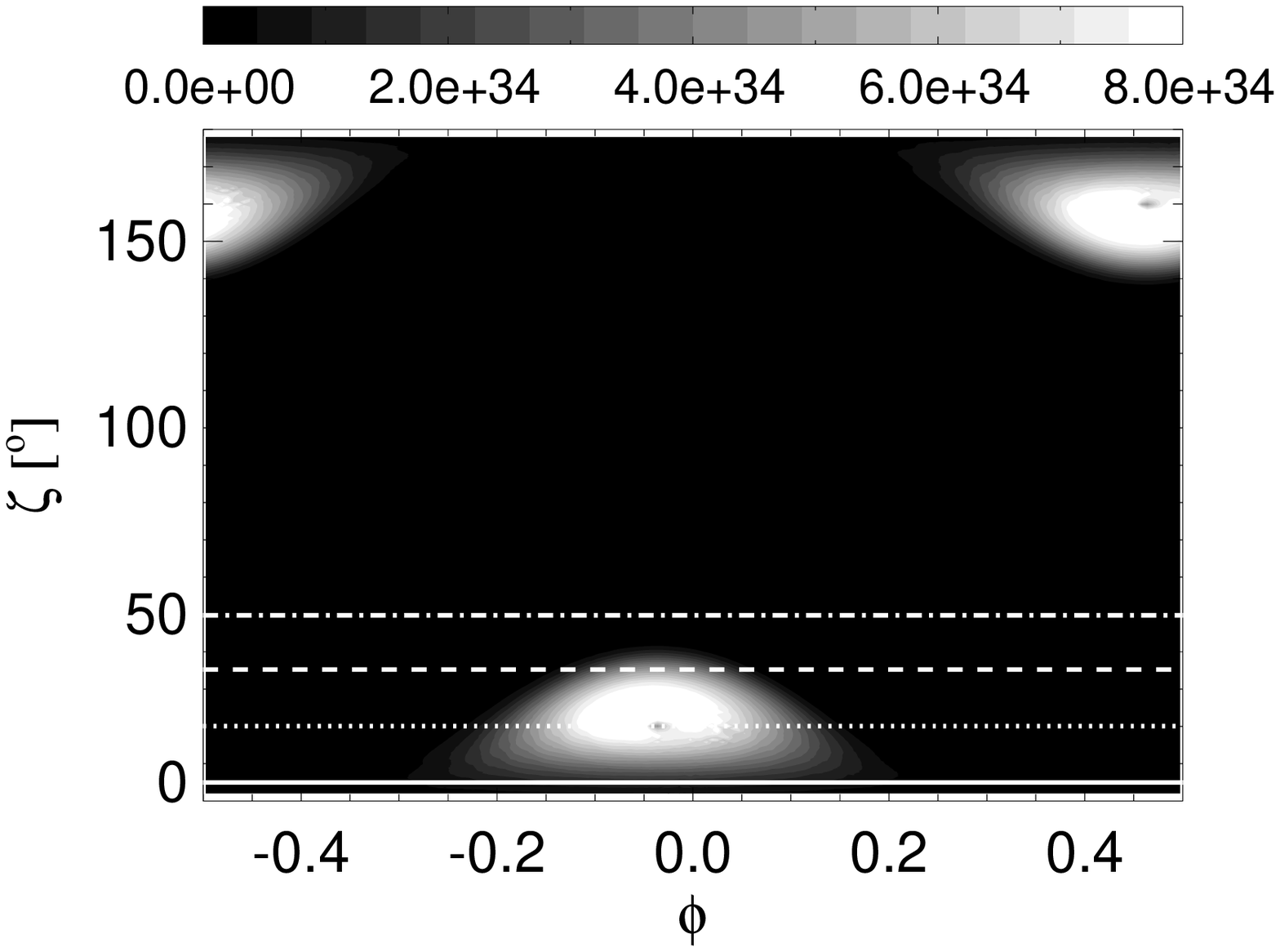} 
        }
    \caption{\small Photon density maps $\dot{N}_{\zeta, \varphi}(E_{\mathrm{min}}, E_{\mathrm{max}})$ in the energy ranges: from 100 MeV to 30 GeV (\textbf{top panel}) and from 30 GeV to 10 TeV (\textbf{bottom panel}). The pulsar rotation phase $\phi = \varphi/2\pi$ is on the x-axis and the viewing angle $\zeta$ in degrees is on the y-axis. White horizontal lines indicate observers for which corresponding light curves have been plotted in Fig.~\ref{profile}. $\zeta = 5^{\circ}, 20^{\circ}, 35^{\circ}, 50^{\circ}$ are depicted with solid, dotted, dashed and dot-dashed line, respectively. See discussion of the photon density maps in the text.}\label{maps}

\end{figure}

We present the results for a single millisecond pulsar in the form of photon density maps, which demonstrate the directional dependence of the photon emission in a chosen energy band. Such map is a contour plot where the photon density defined as 
\begin{equation}
\dot{N}_{\zeta, \varphi}(E_{\mathrm{min}}, E_{\mathrm{max}}) = \int_{E_{\mathrm{min}}}^{E_{\mathrm{max}}}\frac{dN}{dt\: dE\: d\zeta\: d\varphi}dE
\end{equation}
 is coded in grey scale ($N$ is a number of photons). It is presented in the coordinate system where the pulsar rotation phase $\phi = \varphi/2\pi$ is on the x-axis and the observing angle $\zeta$ is on the y-axis. Note that these maps are presented in a different manner than the photon density maps in e.g. \citet{dr03}, \citet{fr05} or \citet{h07}.

\begin{figure}
\hbox{\hspace{0.8cm} $100$ MeV $\leq E \leq 30$ GeV \hspace{3.2cm} $30$ GeV $\leq E \leq 10$ TeV} 
\vspace{0.2cm}
\hbox{ 
    \hspace{-0.7cm}\psfig{figure=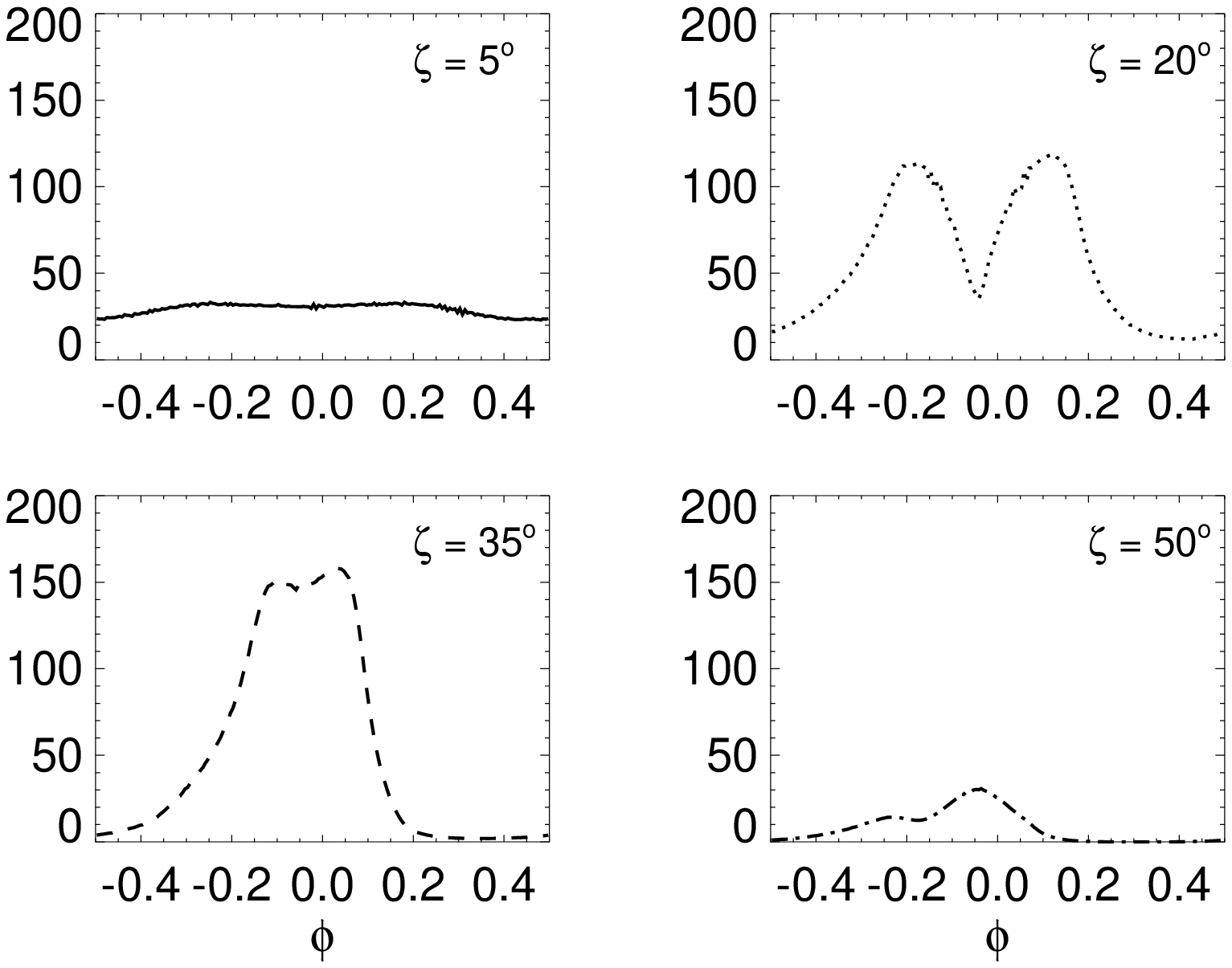,width=6.7cm}
 \hspace{0.5cm}
    \psfig{figure=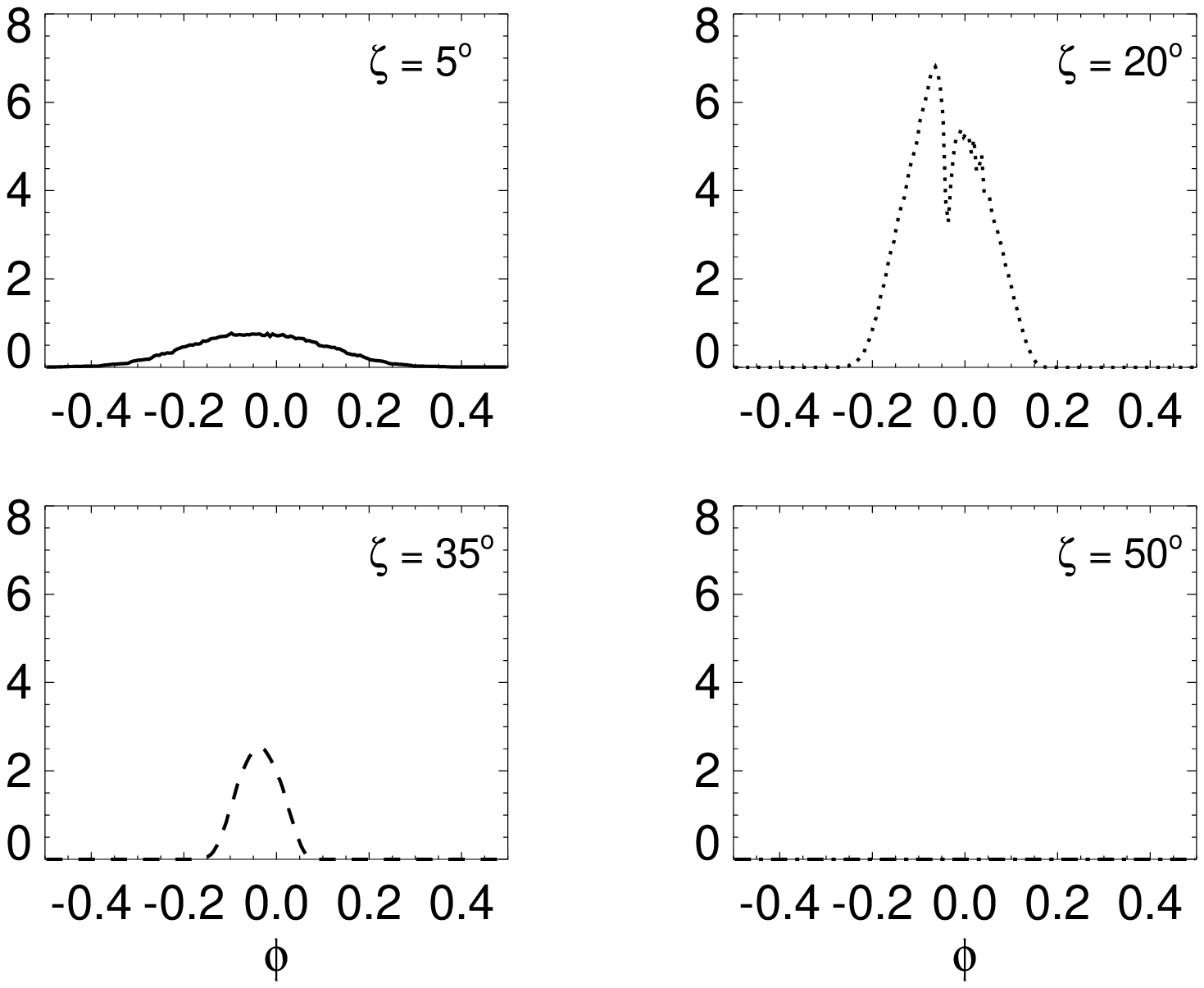,width=6.7cm}
     }
    \caption{\small Pulsar light curves for the observers with $\zeta = 5^{\circ}, 20^{\circ}, 35^{\circ}, 50^{\circ}$ in the energy range from 100 MeV to 30GeV (\textbf{left panel}) and from 30 GeV up to 10 TeV (\textbf{right panel}). The light curves are due to the horizontal cuts through the photon density maps $\dot{N}_{\zeta, \varphi}(E_{\mathrm{min}}, E_{\mathrm{max}})$ presented in Fig.~\ref{maps}. The values of the photon density are in arbitrary units.}\label{profile}
\end{figure}

Two energy bands have been chosen in which the photon density maps of the modelled millisecond pulsar are presented (Fig.~\ref{maps}). The low energy band stretches from 100 MeV to 30 GeV, while the high energy band covers the energies from 30 GeV up to 10 TeV. The energy division corresponds to the energy bands of Fermi GST (the low energy band) and H.E.S.S. II (the high energy band).
Photons with energies below 30 GeV (Fig.~\ref{maps}, top panel) are emitted in a range of the viewing angle $\zeta$ that is wider in comparison with the range for the high energy band (Fig.~\ref{maps}, bottom panel). The emission cone of the high energy photons is centred at the pulsar magnetic axis.
This centring of emission is also traceable in the horizontal cuts made through the photon density maps for chosen observers (Fig.~\ref{profile}).
In the energies above 30 GeV the level of the photon density drops dramatically if the observer moves away from the magnetic axis (Fig.~\ref{profile}, right panel). The decrease in the photon density level is also present in the lower energy band (Fig.~\ref{profile}, left panel); however, it is not so rapid. This behaviour is clearly visible if one compares the light curves for $\zeta = 5^{\circ}$, $20^{\circ}$ and $35^{\circ}$ in different energies. The overall photon density level is lower for the high energy photons.

To have an idea how accurate the estimations of the average power radiated by the pulsar are, we have calculated a pseudo luminosity $L_{\mathrm{pseudo}}$ for both energy ranges. $L_{\mathrm{pseudo}}$ emulates the power inferred by an observer viewing the pulsar at an angle $\zeta$. The emitted radiation is confined to a solid angle $\Omega_{\mathrm{pseudo}} = 1$~sr. The formula for the pseudo luminosity in the energy range from $E_{\mathrm{min}}$ to $E_{\mathrm{max}}$ for a viewing angle $\zeta$ is thus given as 
\begin{equation}
L_{\mathrm{pseudo}}(\zeta; E_{\mathrm{min}}, E_{\mathrm{max}}) = \frac{\Omega_{\mathrm{pseudo}}}{2\pi\;\sin\zeta} 
\int_{E_{\mathrm{min}}}^{E_{\mathrm{max}}}\int_{0}^{2\pi} E\; \frac{dN}{dt\: dE\: d\zeta\: d\varphi}\; d\varphi\; dE \quad.
\label{pseudo}
\end{equation}
$L_{\mathrm{pseudo}}$ is a theoretical counterpart of $L_{\mathrm{obs}}$ inferred from observations in the given energy range $(E_{\mathrm{min}}, E_{\mathrm{max}})$, which can be described with the following formula:
\begin{equation}
L_{\mathrm{obs}}(E_{\mathrm{min}}, E_{\mathrm{max}}) = \Omega_{\mathrm{arb}} \cdot D^{2} \cdot \langle F(E_{\mathrm{min}}, E_{\mathrm{max}})\rangle \quad,
\label{obs}
\end{equation}
where $\Omega_{\mathrm{arb}}$ is a solid angle chosen arbitrarily (1 steradian or sometimes $2\pi$ steradians), $D$ is a distance to the object and $\langle F(E_{\mathrm{min}}, E_{\mathrm{max}})\rangle$ is a phase-averaged energy flux within $(E_{\mathrm{min}}, E_{\mathrm{max}})$ inferred from observations. This formula is frequently used to asses the pulsar luminosity from the observed phase-averaged flux and the known distance to the object. The modelled pseudo luminosities $L_{\mathrm{pseudo}}$ will therefore test the accuracy of the observed luminosities $L_{\mathrm{obs}}$ versus the actual luminosity $L_{\mathrm{actual}}$.

\begin{figure} 
\hspace{0.2cm} \vbox{
\psfig{figure=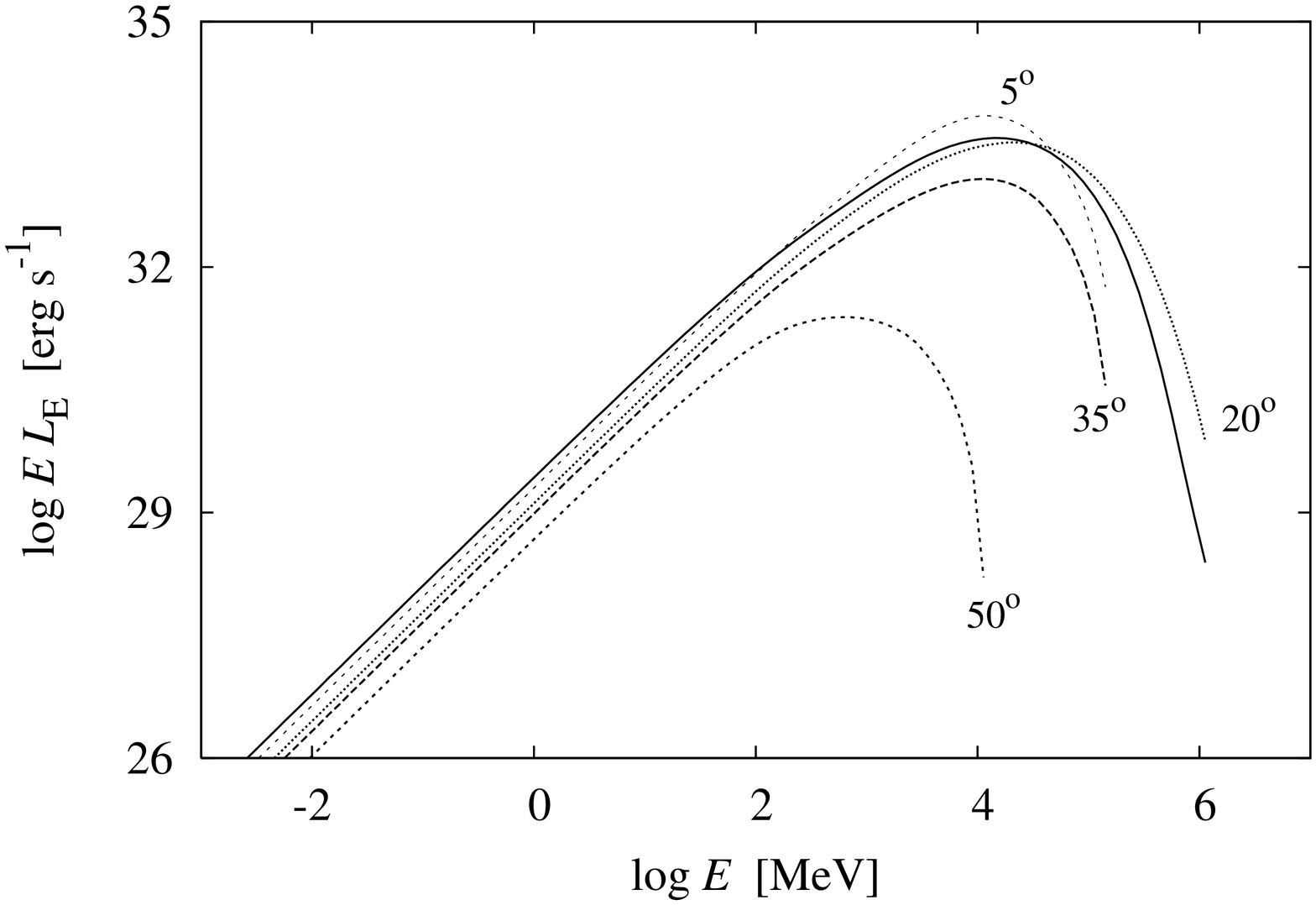,angle=-90,width=12cm}
\psfig{figure=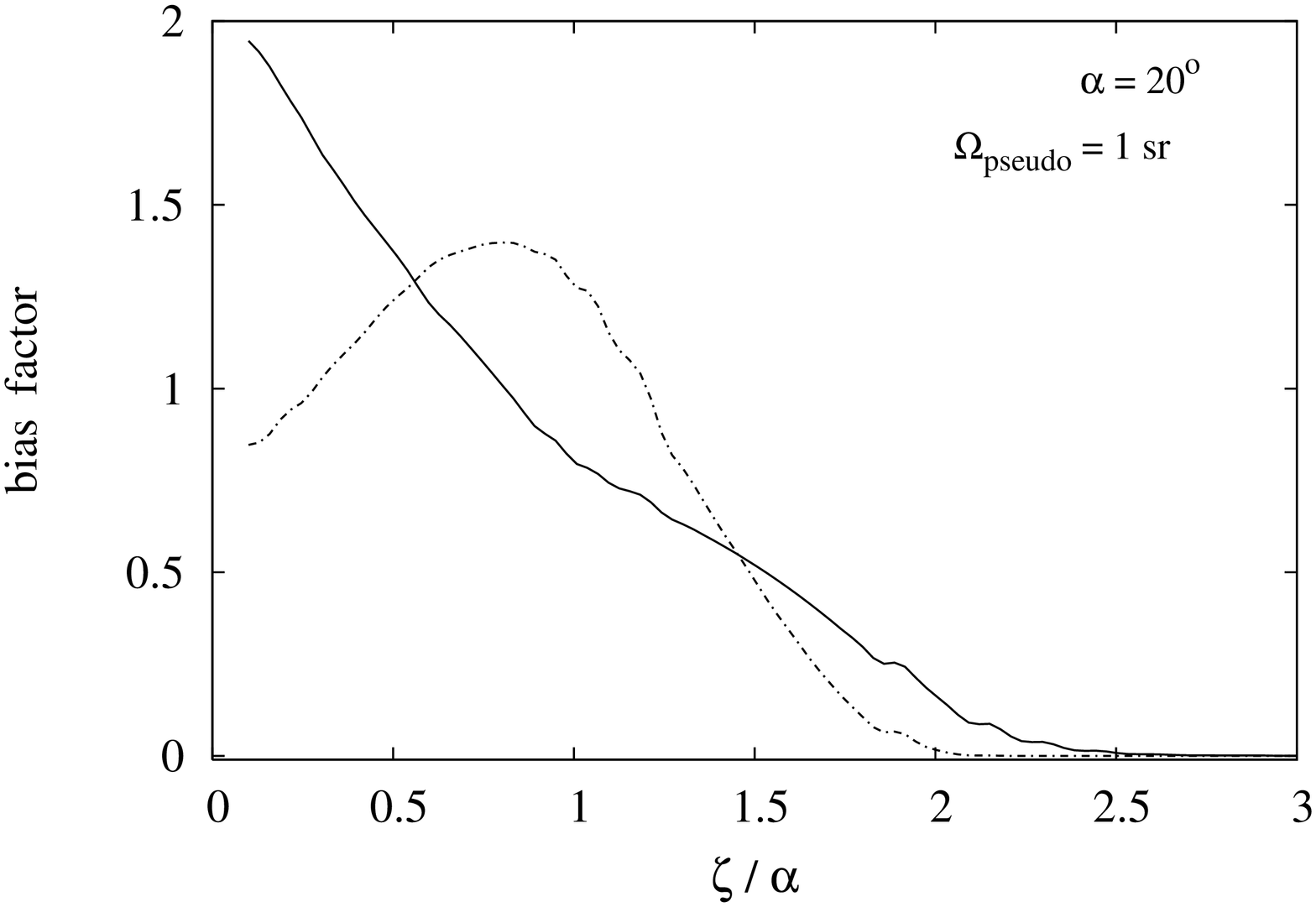,angle=-90,width=12cm}
}
    \caption{\small \textbf{Top panel:} Pseudo luminosity spectra $L_{\mathrm{E, pseudo}; \zeta}$ for the observers with $\zeta = 5^{\circ}, 20^{\circ}, 35^{\circ}, 50^{\circ}$ are shown. Actual output from each pole $L_{\mathrm{E, actual}}$ is presented here with solid line. Here $L_{\mathrm{E}} = \frac{dL}{dE}$; \hspace{0.3cm} $L_{\mathrm{E, pseudo}; \zeta} = \frac{\Omega_{\mathrm{pseudo}}}{2\pi\;\sin\zeta} 
\int_{0}^{2\pi} E\; \frac{dN}{dt\: dE\: d\zeta\: d\varphi}\; d\varphi$ \hspace{0.3cm} and \hspace{0.3cm}
$L_{\mathrm{E, actual}} = \int_{0}^{2\pi} \int_{0}^{\frac{\pi}{2}} E\; \frac{dN}{dt\: dE\: d\zeta\: d\varphi}\; d\zeta\; d\varphi$.
\newline \textbf{Bottom panel:} The dependence of the bias factor $b$ on the viewing angle $\zeta$ scaled with the inclination angle $\alpha$ is presented in the energy range from 100 MeV to 30 GeV and from 30 GeV to 10 TeV with a solid and a dot-dashed line, respectively. See discussion of the bias factor in the text.}\label{pseudolum}
\end{figure}

The theoretical pseudo luminosity spectra are presented in the top panel of Fig.~\ref{pseudolum}. In the same figure the spectrum of the actual power radiated by the pulsar from one pole is shown with solid line.

For each observer we can construct \textit{the bias factor}
\begin{equation}
b \equiv \frac{L_{\mathrm{pseudo}}}{L_{\mathrm{actual}}} \quad,
\label{bf}
\end{equation}
where $L_{\mathrm{pseudo}}$ is given by Eq.~\ref{pseudo} and $L_{\mathrm{actual}}$ is defined as
\begin{equation}
L_{\mathrm{actual}}(E_{\mathrm{min}}, E_{\mathrm{max}}) = \int_{E_{\mathrm{min}}}^{E_{\mathrm{max}}}\int_{0}^{2\pi} \int_{0}^{\frac{\pi}{2}} E\; \frac{dN}{dt\: dE\: d\zeta\: d\varphi}\; d\zeta\; d\varphi\; dE \quad.
\label{actual}
\end{equation}
Both $L_{\mathrm{pseudo}}$ and $L_{\mathrm{actual}}$ are calculated within the same energy range ($E_{\mathrm{min}}$, $E_{\mathrm{max}}$).
The dependence of the bias factor on the viewing angle is presented in the bottom panel of Fig.~\ref{pseudolum}. In the low energy range (solid line) the bias factor equals 1 for $\zeta \simeq 0.8\, \alpha$. The observers with $\zeta \lesssim 0.8\, \alpha$ tend to overestimate the pulsar actual luminosity (Eq.~\ref{actual}) using the pseudo luminosity (Eq.~\ref{pseudo}), while observers with $\zeta \gtrsim 0.8\, \alpha$ underestimate $L_{\mathrm{actual}}$. In the high energy range (dot-dashed line) the situation is different due to the opening angle of radiation being smaller than the inclination angle $\alpha$ and $b = 1$ for $\zeta \simeq 0.25\, \alpha$ and $\zeta \simeq 1.2\, \alpha$. Viewing the pulsar with $\zeta$ in the range from $0.25\, \alpha$ to $1.2\, \alpha$ results in overestimation of $L_{\mathrm{actual}}$. However, the overestimation is rather insignificant (\textit{b} is less than 1.5 in maximum). On the other hand, viewing it with $\zeta$ lower than $0.25\, \alpha$ or higher than $1.2\, \alpha$ leads to underestimation of the pulsar actual luminosity.

Finally, we calculate the gamma-ray efficiency factor\symbolfootnote[2]{$L_{\mathrm{actual}}$ (Eq.~\ref{actual}) is the power output only from one pole. Thus, in order to compare it with $L_{\mathrm{sd}}$ we have to multiply it by 2 to include the output from both poles.} :
\begin{equation}
 \eta_{\gamma}=\frac{2L_{\mathrm{actual}}}{L_{\mathrm{sd}}} \quad.
\end{equation}
Here, $L_{\mathrm{actual}}$ (Eq.~\ref{actual}) is calculated within the energy range stretching from 100 MeV up to 10 TeV.
This factor demonstrates what part of the pulsar spin-down luminosity $L_{\mathrm{sd}}$ is converted into the gamma radiation. The gamma-ray efficiency factor for the modelled millisecond pulsar is $\eta_{\gamma} \simeq 0.05$. This result is in good agreement with $\eta_{\gamma}$ estimated by \citet{hmz02}.

\section{Conclusions}\label{sec4}
The preliminary results of the numerical modelling of millisecond pulsars have been presented in this paper. We have calculated the photon density maps in two energy regimes corresponding to the operational energy bands of Fermi GST and H.E.S.S. II. We have also computed the pseudo luminosities for different viewing angles and on their basis the plot presenting the behaviour of the bias factor has been constructed.
From the comparison of the bias factor - viewing angle dependence in different energy bands we may infer information on accuracy of the so-called pseudo luminosity $L_{\mathrm{pseudo}}$ as an estimation of the pulsar actual luminosity $L_{\mathrm{actual}}$ in $\gamma$-rays. It turns out that there are very few favourable $\zeta$ for which $L_{\mathrm{pseudo}}$ is a good estimate of $L_{\mathrm{actual}}$. Most often $L_{\mathrm{actual}}$ tends to be under/overestimated.
 The gamma-ray efficiency factor for the modelled millisecond pulsar has been calculated. Its value $\eta_{\gamma} \simeq 0.05$ is in agreement with the values in \citet{hmz02}. However, further work concerning modelling, especially the inclusion of radiative processes like synchrotron radiation and inverse Compton scattering important for pulsar X-ray emission, has to be done in order to draw accurate conclusions about millisecond pulsars. Also simulations for other sets of pulsar parameters, especially for different inclination angles $\alpha$, have to be carried out.

We intend to use the results of modelling as a tool for predicting the detectability of millisecond pulsars with high energy observatories like Fermi GST, H.E.S.S. II and even CTA \citep{fr05}. Furthermore, we plan to develop the population synthesis studies of high energy pulsars that would incorporate the results of the modelling described briefly in this paper.

\section*{Acknowledgements}
AZ wants to thank Bronek Rudak and Jarek Dyks for many helpful discussions and critical remarks on the paper. AZ is also grateful to the organisers for giving her the opportunity to attend \textit{the LAPIS School} and for financial support.
This work was supported by the grant N203 017 31/2872 of the Ministry of Science and Higher Education and the Polish Astroparticle Network 621/E-78/SN-0068/2007.

\end{document}